\documentclass[smallextended,twocolumn]{svjour3}
\smartqed
\PassOptionsToPackage{numbers,sort&compress,...}{natbib}
\usepackage{graphicx}
\usepackage{graphics}
\usepackage{color}
\usepackage{framed}
\usepackage{lipsum}
\usepackage{citesort}
\usepackage{amssymb,amsmath,amsfonts}
\makeatletter
\newcommand*{\gnuplotinput}[2][1.0]{%
	\begingroup
	\let\@gnplt@input@includegraphics=\includegraphics
	\def\includegraphics##1{\@gnplt@input@includegraphics[scale=#1]{#2}}%
	\let\@gnplt@input@picture=\picture
	\def\picture{\unitlength=#1\unitlength\relax\@gnplt@input@picture}%
	\input{#2}%
	\endgroup
}
\makeatother
\usepackage{tikz}

\newcommand{\formula}[3]{C$_{#1}$H$_{#2}$O$_{#3}$}

\begin{document}

\title{Evaluation of Peppermint Leaf Flavonoids as SARS-CoV-2 Spike Receptor-Binding Domain Attachment Inhibitors to the Human ACE2 Receptor: A Molecular Docking Study}

\author{M. L. Pereira J\'unior$^{a,}$* \and R. T. de Sousa Junior$^b$ \and G. D. Amvame Nze$^b$ \and W. F. Giozza$^b$ \and L. A. Ribeiro J\'unior$^a$}
\institute{
$^a$Institute of Physics, University of Bras\'{i}lia, P.O. Box 04531, 70.919-970, Bras\'{i}lia, DF, Brazil. \\$^b$Department of Electrical Engineering, University of Bras\'{i}lia, 70.919-970, Bras\'{i}lia, DF, Brazil. \\ *Corresponding Author: \email{marcelolpjunior@gmail.com}
}

\date{Received: xx/xx/2021 / Accepted: xx/xx/2021}

\maketitle

\begin{abstract}
Virtual screening is a computational technique widely used for identifying small molecules which are most likely to bind to a protein target. Here, we performed a molecular docking study to propose potential candidates to prevent the RBD/ACE2 attachment. These candidates are sixteen different flavonoids present in the peppermint leaf. Results showed that Luteolin 7-O-neohesperidoside is the peppermint flavonoid with a higher binding affinity regarding the RBD/ACE2 complex (about -9.18 Kcal/mol). On the other hand, Sakuranetin presented the lowest affinity (about -6.38 Kcal/mol). Binding affinities of the other peppermint flavonoids ranged from -6.44 Kcal/mol up to -9.05 Kcal/mol. The binding site surface analysis showed pocket-like regions on the RBD/ACE2 complex that yield several interactions (mostly hydrogen bonds) between the flavonoid and the amino acid residues of the proteins. This study can open channels for the understanding of the roles of flavonoids against COVID-19 infection.  
\end{abstract}


\section{Introduction}

The COVID-19 is an infectious disease caused by the coronavirus SARS-CoV-2 \cite{chen_TL,rothan_JA,guo_MMR,tu_IJMS,wang_IJAA}. It has reached the status of a pandemic in March of 2020. Up to January of 2021, it has already infected more than 100 million people, leading to the death of more than 2 million ones \cite{oms_url}. Since the earlier stages of this pandemic, a worldwide effort has been devoted to producing vaccines and antiviral drugs to combat this virus. Some successful investigations yielded vaccines that have started to be applied very recently \cite{sharma_FiI,voysey_TheLancet,corbett_NEJM,baden_NEJM,anderson_NEJM,zhang_medrxiv,jones_TheLancet,burki_TheLancet,jones_TheLacent}. Despite the beginning of vaccination, no consensus about an efficient treatment for already infected patients has been reached so far. 

Sars-CoV-2 has a crown-like (spherical) form, and its surface protein (Spike) is directly involved in the infectious process \cite{wu_BBaI,chhikara_CBL,liang_FiI}. The receptor of this virus in human cells is the angiotensin-converting enzyme 2 (ACE2) \cite{zhang2020angiotensin,li2003angiotensin,kuba2005crucial}. Sars-CoV-2 surface protein has two subdivisions named S1 and S2, being S1 the receptor-binding domain (RBD) \cite{xiu2020inhibitors,tai2020characterization,spinello2020rigidity,han2020computational}. The RBD plays a major role in the attachment mechanism of Spike protein to ACE2 \cite{singh2020serine}. After the attachment between them, the virus enters the cell and starts the replication process \cite{xiu2020inhibitors}. In this sense, the strategy of virtual screening for possible inhibitors for the RBD/ACE2 attachment \cite{basu2020molecular} may pave the way for novel therapeutic approaches for the treatment of COVID-19.

Drug repurposing is a feasible way to combat diseases with some similarities \cite{zhou2020network,gordon2020sars,pandey2020potential}. In this scenario, the use of phytochemicals is always an important option to be considered \cite{swain2020phytochemicals}. Among their sub-classes, the flavonoids --- a class of small molecules found in fruits, vegetables, flowers, honey, teas, and wines --- stand out \cite{panche2016flavonoids,harborne2013flavonoids,hertog1992content}. Their pharmacological properties include antimicrobial, antioxidant, anti-inflammatory, and antiviral functions \cite{kumar2013chemistry,sharma2006pharmacological,kim2004anti}. 

Flavonoids have been employed as inhibitors for the infection mechanism of several diseases \cite{cushnie2005antimicrobial}. Among them, one can mention malaria, leishmaniasis, Chagas, and dengue \cite{lehane2008common,muzitano2009oral,marin2011vitro,sanchez2000antiviral,kiat2006inhibitory,de2015flavonoids}. They have also been considered in studies aimed at developing therapeutic approaches for cancer treatment \cite{abotaleb2019flavonoids,wang2000therapeutic,eunjung2012}. Very recently, it was reported that Luteolin (a flavonoid found in leaves and shells) is efficient as an anti-inflammatory that can interact with the Sars-CoV-2 surface \cite{yu2020computational} and its main protease \cite{muchtaridi2020natural}. More specifically, it is adsorbed in the Spike protein, inhibiting the Sars-CoV-2 attachment to the ACE2, thus preventing infection. Ngwa and colleagues used computer simulations to address the feasibility of Caflanone, Hesperetin, and Myricetin flavonoids in acting as inhibitors for the ACE2 active site attachment \cite{ngwa_M}. Their results pointed to the ability of Caflanone in inhibiting the transmission of the Sars-CoV-2 virus from mother to fetus in pregnancy. Pandey \textit{et. al.} conducted molecular docking and dynamics simulations considering ten flavonoid and non-flavonoid compounds (by using phytochemicals and hydroxychloroquine, respectively) to verify their performance in inhibiting the RBD/ACE2 interaction \cite{pandey_JBSD}. Their findings indicate that Fisetin, Quercetin, and Kamferol molecules couple to RBD/ACE2 complex with good binding affinities. In this sense, they can be explored as possible anti-Sars-CoV-2 agents. Despite the success of these molecules inhibiting the RBD/ACE2, other flavonoids should be tested to broaden the list of possible inhibitors and to confirm their potential in developing new therapeutic approaches for the treatment of COVID-19.      

Herein, in silico molecular docking analysis was carried out to propose potential flavonoid candidates in preventing the RBD/ACE2 attachment. These candidates are sixteen different flavonoids present in the Peppermint (Mentha piperita) leaf \cite{DOLZHENKO201067,McKayAreviewof,bodalska2020analysis,areias2001phenolic,riachi2015peppermint,mahendran2020ethnomedicinal,peterson1998flavonoids}. Peppermint is a perennial herb and medicinal plant native to Europe widely used for treating stomach pains, headaches, and inflammation of muscles \cite{McKayAreviewof,mahendran2020ethnomedicinal,peterson1998flavonoids}. Well-known for their flavoring and fragrance traits, peppermint leaves and the essential oil extracted from them are used in food, cosmetic and pharmaceutical products \cite{McKayAreviewof,bodalska2020analysis,DOLZHENKO201067,areias2001phenolic}. Our results revealed that Luteolin 7-O-neohesperidoside is the peppermint flavonoid with a higher binding affinity regarding the RBD/ACE2 complex (about -9.18 Kcal/mol). On the other hand, Sakuranetin was the one with the lowest affinity (about -6.38 Kcal/mol). Binding affinities of the other peppermint flavonoids ranged from -6.44 Kcal/mol up to -9.05 Kcal/mol. These binding affinities are equivalent to other ones reported in the literature for the interaction between flavonoids and the RBD/ACE2 complex \cite{DeOliveira2020,Muhseen2020,Hu2020,muchtaridi2020natural,Basu2020,Istifli2020,Peterson2020,yu2020computational,dourav_JBSD,RUSSO2020109211}. Moreover, the binding site surface analysis showed pocket-like regions on the RBD/ACE2 complex that yield several interactions (mostly hydrogen bonds) between the flavonoid and the amino acid residues of the proteins. Definitively, experimental studies and clinical trials should be further performed to evaluate the efficacy of these compounds in the inhibition of the RBD/ACE2 attachment. 

\section{Materials and Methods}   

\begin{figure*}[htb]
	\centering
	\includegraphics[width=1.0\linewidth]{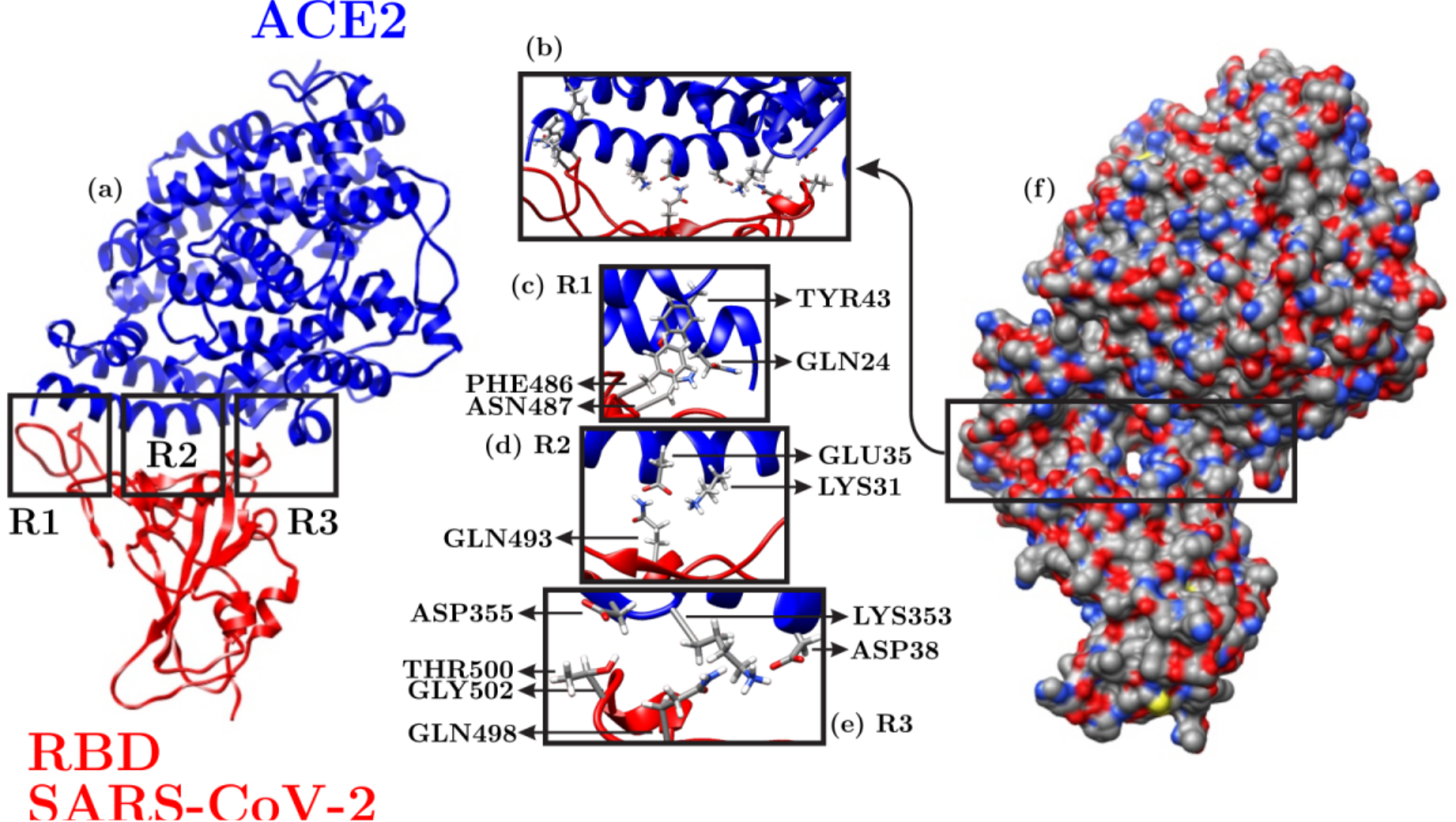}
	\caption{Schematic representation of the (a) main proteins involved RBD/ACE2 interaction. These proteins were obtained from Protein Data Bank, ID 6M0J \cite{lan2020structure}. (b) The binding site surface has the following color scheme: gray, red, blue, and white for carbon, oxygen, nitrogen, and hydrogen atoms, respectively. Only the three regions (R1, R2, and R3) were considered in the docking processes since they define the whole RBD/ACE2 interface. The TYR4, GLN24, PHE486, and ASN487 are the residues present in the region R1; GLU35, LYS31, and GLN493 are the residues present in the region R2; ASP355, THR500, GLY502, GLN498, LYS353, and ASP38 are the residues present in the region R3.}
	\label{fig:main_protein}
\end{figure*}

Since Sars-CoV-2 infects human cells through the RBD/ACE2 coupling, the idea of checking for small molecules that may inhibit this interaction is recurring and can be useful to propose a combatant drug \cite{Bojadzic2020.10.22.351056}. Here, we used molecular docking to study the interaction between the peppermint flavonoids with the RBD/ACE2 complex. Below, we present the proteins, inhibitors (flavonoids), and the computational protocol involved in our study.

\subsection{Protein Preparation}

Figure \ref{fig:main_protein} presents the main proteins involved RBD/ACE2 interaction that were obtained from Protein Data Bank, ID 6M0J \cite{lan2020structure}. In the left panel of this figure, the ACE2 protein is in blue, while the RBD Sars-CoV-2 one is in red. Three essential regions of inhibition between these proteins were highlighted with the black squares R1, R2, and R3. In the right side of Figure \ref{fig:main_protein} we show the binding site surface colored as gray, red, blue, and white for carbon, oxygen, nitrogen, and hydrogen atoms, respectively. The yellow rectangle highlights the total surface for inhibition with a clear cavity within region R2. The protein resolution is 2.45 \r{A}, and no pKa prediction was carried out. The modeled structure has 41 residues less than the deposited one, but all the important residues in the RDB/ACE2 interface were considered in our study. Just metal ions were considered in the docking study, water molecules were not included. 

\subsection{Ligand Preparation}

\begin{figure*}[!htb]
	\centering
	\includegraphics[width=0.8\linewidth]{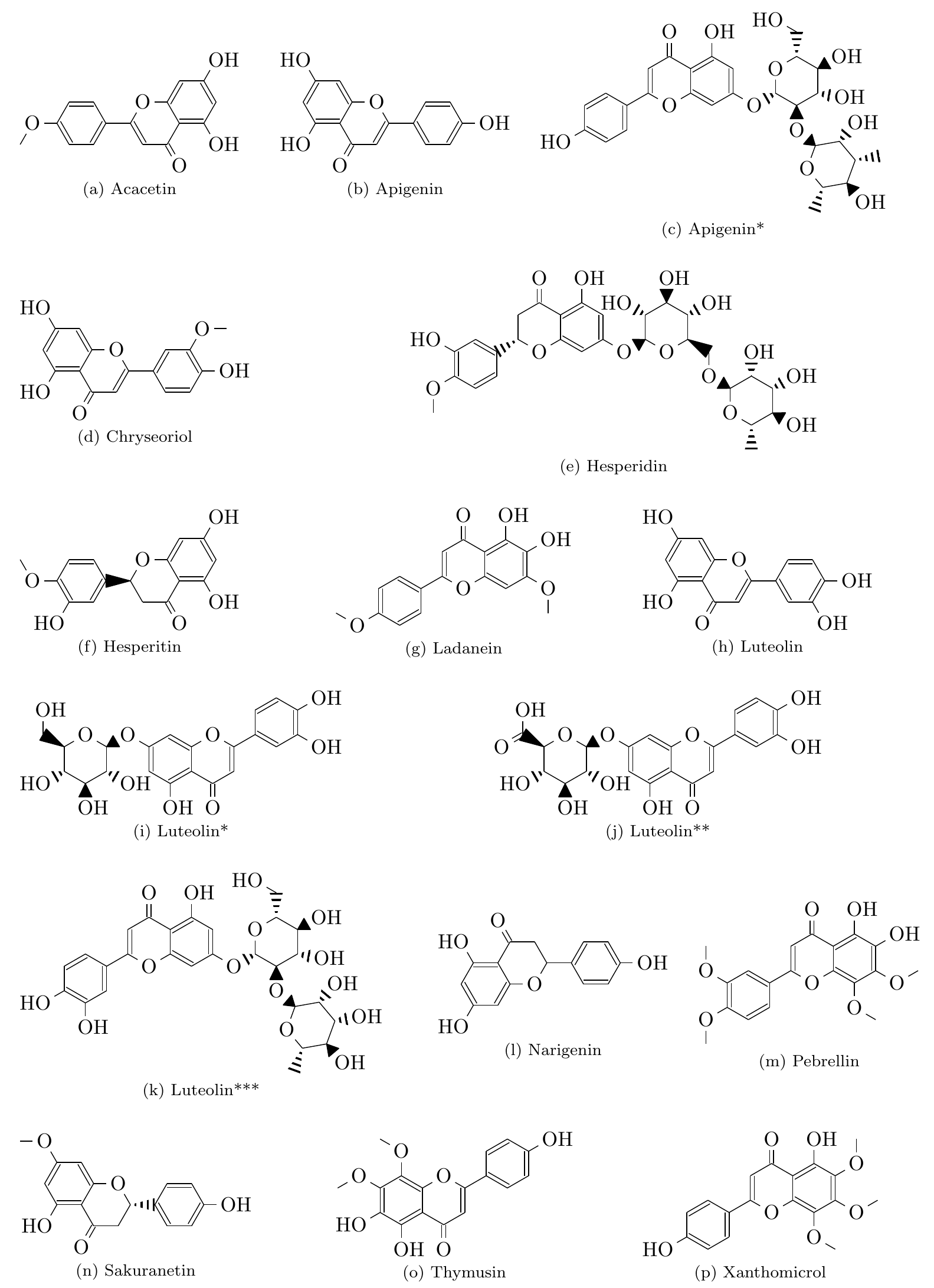}
	\caption{Chemical structure of peppermint leaf flavonoids: (a )Acacetin, (b) Apigenin, (c) Apigenin 7-O-neohesperidoside (Apigenin*), (d) Chryseoriol, (e) Hesperidin, (f) Hesperitin, (g) Ladanein, (h) Luteolin, (i) Luteolin 7-O-glucoside (Luteolin*), (j) Luteolin 7-O-glucuronide (Luteolin**), (k) Luteolin 7-O-neohesperidoside (Luteolin***), (l) Narigenin, (m) Pebrellin, (n) Sakuranetin, (o) Thymusin, and (p) Xanthomicrol.}
	\label{fig:flavonoids}
\end{figure*}

The peppermint leaf contains sixteen flavonoids \cite{DOLZHENKO201067,areias2001phenolic}, classified into three subcategories: Flavones (Flavonols), Flavorings, and Flavanones \cite{DOLZHENKO201067,areias2001phenolic}. The flavonoids studied here are Acacetin, Apigenin, Apigenin 7-O-neohesperidoside (Apigenin*), Chryseoriol, Hesperidin, Hesperitin, Ladanein, Luteolin, Luteolin 7-O-glucoside (Luteolin*), Luteolin 7-O-glucuronide (Luteolin**), Luteolin 7-O-neohesperidoside (Luteolin***), Narigenin, Pebrellin, Sakuranetin, Thymusin, and Xanthomicrol.  Their 3D structures were extracted from PubChem \cite{pub_url}. The chemical structures of these flavonoids can be seen in figure \ref{fig:flavonoids}, while relevant information such as PubChem ID, molecular weight, molecular formula, and subcategory of the flavonoid is presented in table \ref{tab:lig-id}.

\begin{table*}[htb]
	\centering
	\begin{tabular}{|c|c|c|c|c|}
		\hline
		\textbf{Compound}             & \textbf{PubChem CID} & \textbf{Mol. Weight (g/mol)} & \textbf{Mol. Formula} & \textbf{Type}          \\ \hline
		Acacetin                      & 5280442              & 284.26                            & \formula{16}{12}{5}                  & Flavones and Flavonols \\ \hline
		Apigenin                      & 5280443              & 270.24                            & \formula{15}{10}{5}                   & Flavones and Flavonols \\ \hline
		Apigenin* & 5282150              & 578.5                             & \formula{27}{30}{14}                  & Flavones and Flavonols \\ \hline
		Chryseoriol                   & 5280666              & 300.26                            & \formula{16}{12}{6}                   & Flavones and Flavonols \\ \hline
		Hesperidin                    & 10621                & 610.6                             & \formula{28}{34}{15}                  & Flavorings             \\ \hline
		Hesperitin                    & 72281                & 302.28                            & \formula{16}{14}{6}                   & Flavanones             \\ \hline
		Ladanein                      & 3084066              & 314.29                            & \formula{17}{14}{6}                   & Flavones and Flavonols \\ \hline
		Luteolin                      & 5280445              & 286.24                            & \formula{15}{10}{6}                   & Flavones and Flavonols \\ \hline
		Luteolin*        & 5280637              & 448.4                             & \formula{21}{20}{11}                  & Flavones and Flavonols \\ \hline
		Luteolin**      & 5280601              & 462.4                             & \formula{21}{18}{12}                  & Flavones and Flavonols \\ \hline
		Luteolin*** & 5282152              & 594.5                             & \formula{27}{30}{15}                  & Flavones and Flavonols \\ \hline
		Naringenin                    & 932                  & 272.25                            & \formula{15}{12}{5}                   & Flavorings             \\ \hline
		Pebrellin                     & 632255               & 374.3                             & \formula{19}{18}{8}                   & Flavones and Flavonols \\ \hline
		Sakuranetin                   & 73571                & 286.28                            & \formula{16}{14}{5}                   & Flavanones             \\ \hline
		Thymusin                      & 628895               & 330.29                            & \formula{17}{14}{7}                   & Flavones and Flavonols \\ \hline
		Xanthomicrol                  & 73207                & 344.3                             & \formula{18}{16}{7}                   & Flavones and Flavonols \\ \hline
	\end{tabular}
	\caption{Potential inhibitors (peppermint leaf flavonoids) of RBD/ACE2 complex and their compound information.}
	\label{tab:lig-id}
\end{table*}

\subsection{Molecular Docking Simulation}

Molecular docking consists of computationally analyze the non-covalent binding between macromolecules (receptor) and small molecules (ligand). Here, the macromolecule is the RBD/ACE2 protein complex (Figure \ref{fig:main_protein}), while the ligands are the sixteen flavonoids present in the peppermint leaf (Figure \ref{fig:flavonoids}). SWISSDOCK server was used for the docking simulations \cite{grosdidier2011fast,grosdidier2011swissdock}. In SWISSDOCK, the docking energies are obtained through the CHARMM (Chemistry at HARvard Macromolecular Mechanics) force field \cite{grosdidier2011swissdock,grosdidier2011fast} using a blind docking strategy that spans over 100 trial configurations for each target/ligand input \cite{cheng2020discovery}. The target/ligand configuration with higher binding affinity is selected using the UCFS CHIMERA software \cite{pettersen2004ucsf}, a visualization tool capable of directly import data from the SWISSDOCK server. Finally, the Protein-Ligand Interaction Profiler (PLIP) server \cite{salentin2015plip} is used to characterize the target/ligand interaction for the configuration with a higher binding affinity for each flavonoid regarding the RBD/ACE2 complex. It is worth mentioning that the screening for the ligand position was limited just to the ACE2/RDB interface (regions R1, R2, and R3 in the left panel of Figure \ref{fig:main_protein}). This interface is the crucial region to be considered for blocking the coronavirus entry and replication cycle. The simulation (docking) box used in the screening for the ligand position was limited just to the ACE2/RDB interface (regions R1, R2, and R3 in the left panel of Figure \ref{fig:main_protein}). The docking box has 27.5 \r{A} $\times$ 9.0 \r{A} $\times$ 8.5 \r{A} of dimension and it was centered at (31.5,-36.0,1.5) \r{A}. These parameters cover the three regions depicted in Figure \ref{fig:main_protein}. The accuracy in estimating the ligand positions and related binding affinities are $\pm$2 \r{A} and $\pm$0.01 Kcal/mol, respectively. 

\section{Results}

\begin{table}[htb]
	\centering
	\begin{tabular}{|c|c|}
		\hline
		\textbf{Compound}             & \textbf{$ \Delta G $ {[}Kcal/mol{]}} \\ \hline
		Acacetin                      & -6.70                                \\ \hline
		Apigenin                      & -6.87                                \\ \hline
		Apigenin 7-O-neohesperidoside & -8.08                                \\ \hline
		Chryseoriol                   & -6.78                                \\ \hline
		Hesperidin                    & -8.67                                \\ \hline
		Hesperitin                    & -6.80                                \\ \hline
		Ladanein                      & -6.56                                \\ \hline
		Luteolin                      & -7.24                                \\ \hline
		Luteolin 7-O-glucoside        & -8.01                                \\ \hline
		Luteolin 7-O-glucuronide      & -7.74                                \\ \hline
		Luteolin 7-O-neohesperidoside & -9.18                                \\ \hline
		Naringenin                    & -6.44                                \\ \hline
		Pebrellin                     & -7.07                                \\ \hline
		Sakuranetin                   & -6.38                                \\ \hline
		Thymusin                      & -6.94                                \\ \hline
		Xanthomicrol                  & -6.83                                \\ \hline
	\end{tabular}
	\caption{Peppermint leaf-based flavonoid candidates undergoing docking experiment with their most favorable conformation (lowest binding affinity $\Delta G$ in Kcal/mol).}
	\label{tab:deltaG}
\end{table}

\begin{figure*}[!htb]
	\centering
	\includegraphics[width=0.8\linewidth]{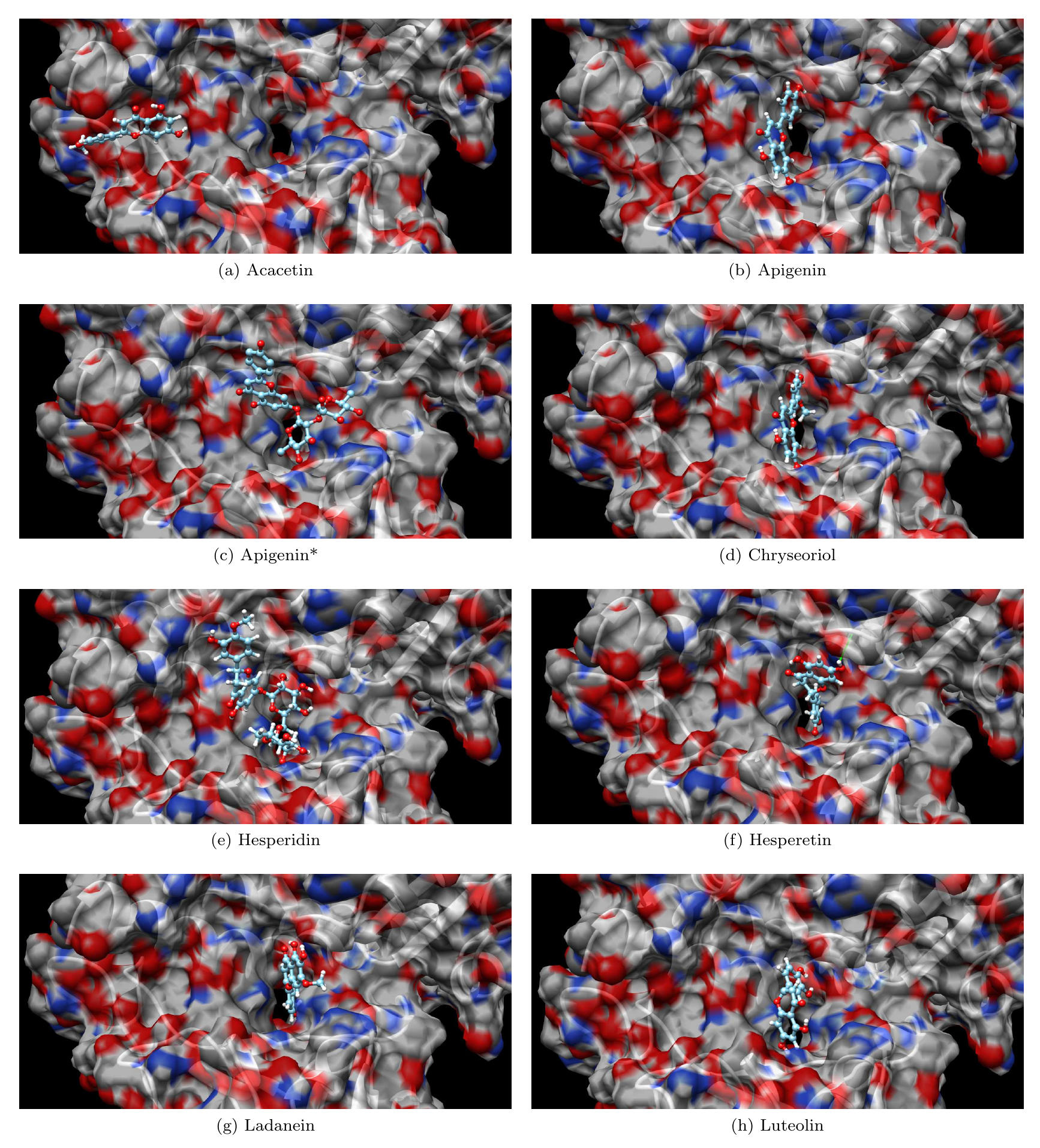}
	\caption{binding site surface (BSS) for the putative best docking target/ligand configurations of (a) Acacetin, (b) Apigenin, (c) Apigenin$^{*}$, (d) Chryseoriol, (e) Hesperidin, (f) Hesperetin, (d) Ladanein, and (d) Luteolin.}
	\label{fig:protein_adsorption-01}
\end{figure*}

\begin{figure*}[!htb]
	\centering
	\includegraphics[width=0.8\linewidth]{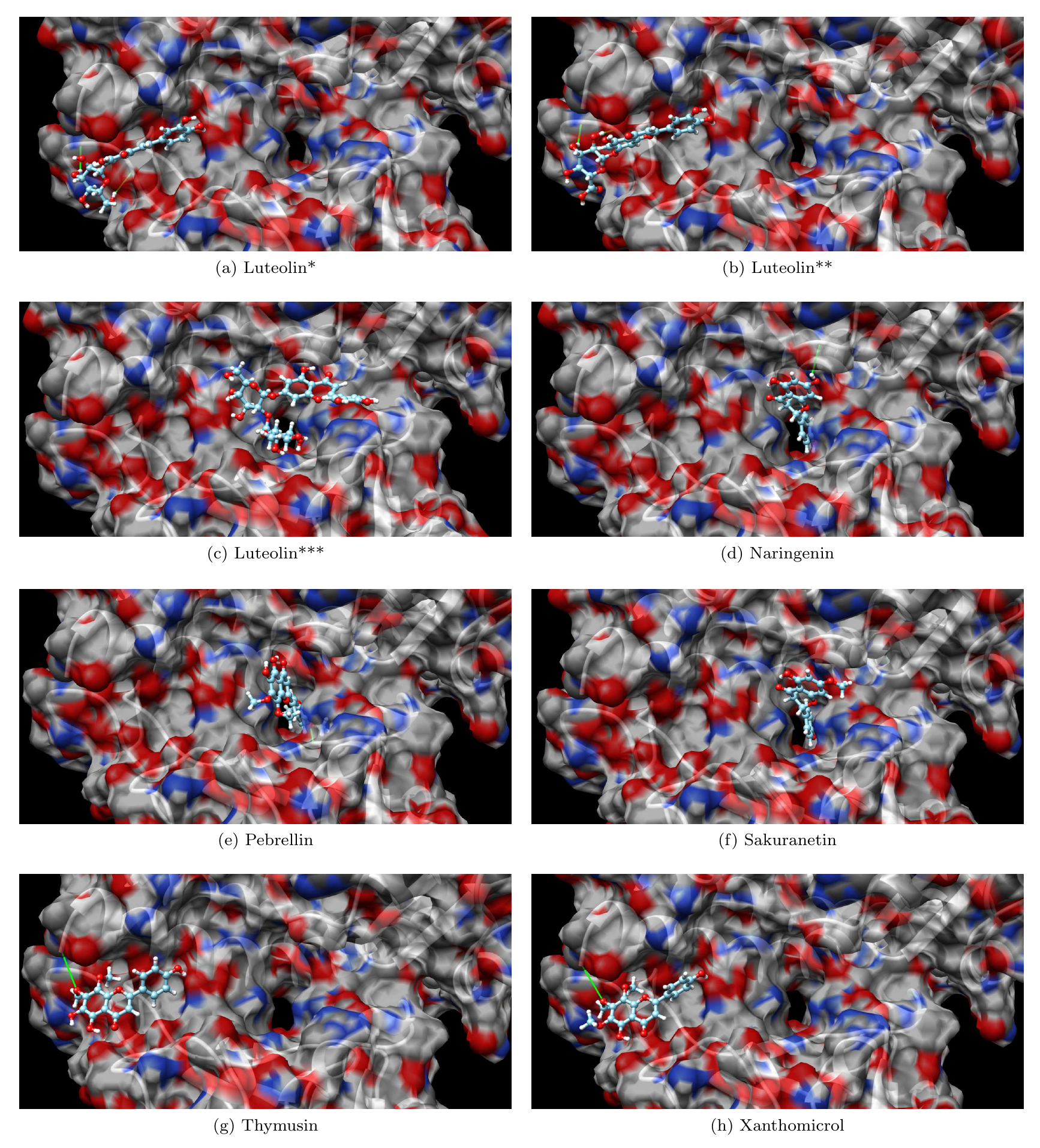}
	\caption{binding site surface (BSS) for the putative best docking target/ligand configurations of (a) Luteloin$^{*}$, (b) Luteloin$^{**}$, (c) Luteloin$^{***}$, (d) Naringenin, (e) Pebrellin, (f) Sakuranetin, (d) Thymusin, and (d) Xanthomicrol.}
	\label{fig:protein_adsorption-02}
\end{figure*}

\begin{figure*}[!htb]
	\centering
	\includegraphics[width=0.8\linewidth]{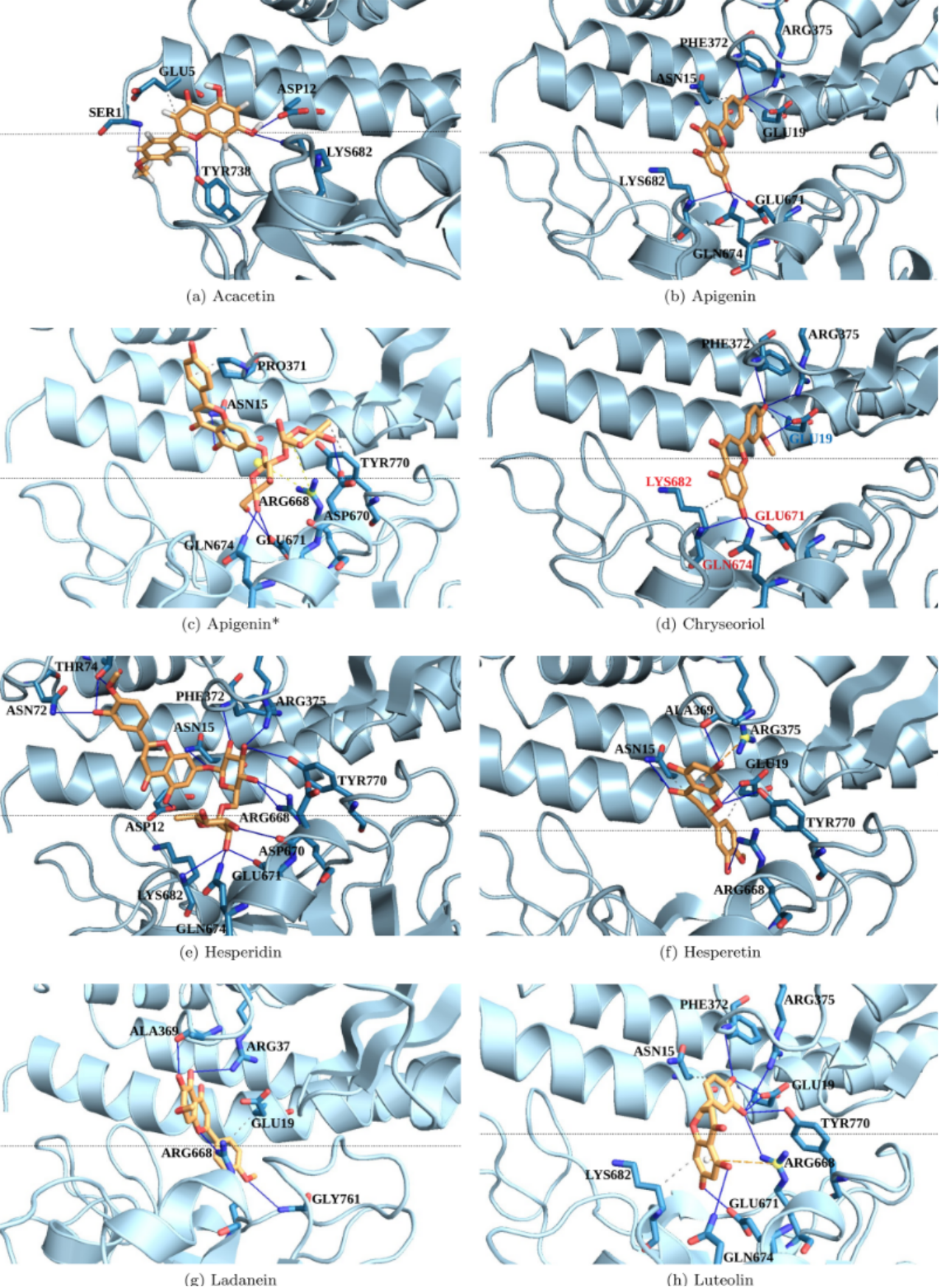}
	\caption{PLIP docked poses for the RBD/ACE2 interaction with (a) Acacetin, (b) Apigenin, (c) Apigenin$^{*}$, (d) Chryseoriol, (e) Hesperidin, (f) Hesperetin, (d) Ladanein, and (d) Luteolin. The hydrogen, hydrophobic, and $\pi$-staking bonds are denoted by the blue, dashed gray, and dashed yellow lines, respectively. The yellow sphere represents the charge center.}
	\label{fig:protein-ligand-01}
\end{figure*}  

\begin{figure*}[!htb]
	\centering
	\includegraphics[width=0.8\linewidth]{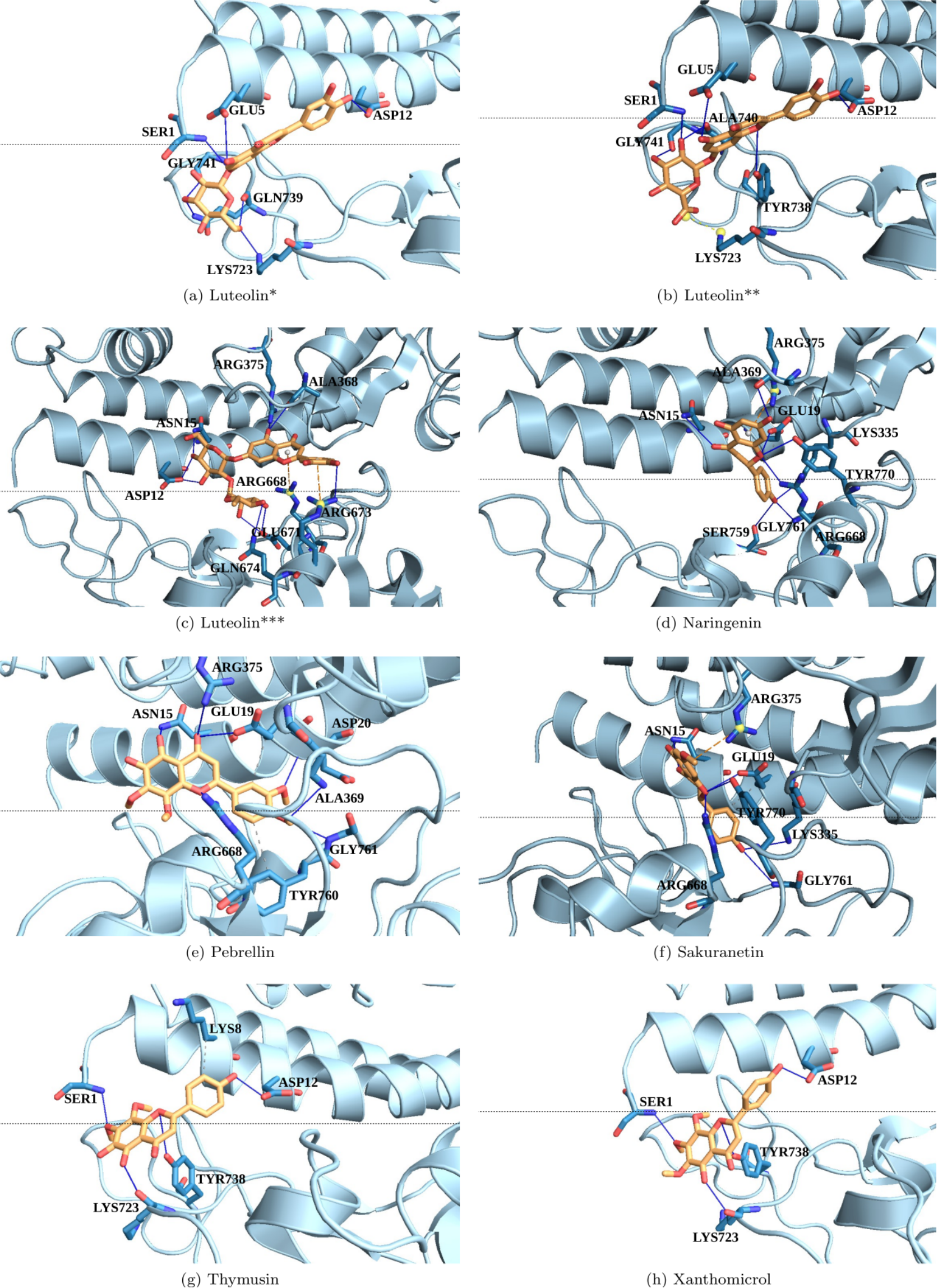}
	\caption{PLIP docked poses for the RBD/ACE2 interaction with (a) Luteloin$^{*}$, (b) Luteloin$^{**}$, (c) Luteloin$^{***}$, (d) Naringenin, (e) Pebrellin, (f) Sakuranetin, (d) Thymusin, and (d) Xanthomicrol. The hydrogen, hydrophobic, and $\pi$-staking bonds are denoted by the blue, dashed gray, and dashed yellow lines, respectively. The yellow sphere represents the charge center. ACE2 and RBD moieties are shown above and below the horizontal line, respectively.}
	\label{fig:protein-ligand-02}
\end{figure*}

After successful docking of the peppermint flavonoids to the RBD/ACE2 complex, several modes of ligand/target interactions were generated with a particular docking score (binding affinity). The binding mode with the lowest binding affinity is regarded as the best one, once it tends to be the most stable. The binding affinity results ($\Delta G$) obtained here are summarized in Table \ref{tab:deltaG}. SWISSDOCK simulations for all the ligands in Figure \ref{fig:flavonoids} revealed significant binding affinities with the target RBD/ACE2 proteins. Luteolin 7-O-neohesperidoside is the peppermint flavonoid with a higher binding affinity regarding the RBD/ACE2 complex (about -9.18 Kcal/mol). On the other hand, Sakuranetin was the one with the lowest affinity (approximately -6.38 Kcal/mol). Binding affinities of the other peppermint flavonoids ranged from -6.44 Kcal/mol up to -9.05 Kcal/mol. As one can note in Tables \ref{tab:lig-id} and \ref{tab:deltaG}, the best docked flavonoids have greater molecular weight. All the binding affinities are close to the ones reported for the RBD/ACE2 interaction with other species of flavonoids \cite{DeOliveira2020,Muhseen2020,Hu2020,muchtaridi2020natural,Basu2020,Istifli2020,Peterson2020,yu2020computational,dourav_JBSD,RUSSO2020109211}. Moreover, they can outperform the binding affinities reported by docking studies using other types of compounds targeting RBD/ACE2 \cite{Basu2020,luan2020targeting,alexpandi2020quinolines,choudhary2020identification,trezza2020integrated,wei2020silico,panda2020structure,utomo2020revealing}, such as Chloroquine and Hydroxychloroquine, which are lower than -8.0 Kcal/mol \cite{Basu2020}. This fact can be attributed to the abundant phenolic hydroxyl group in flavonoids. The hydroxyl group in the sugar group of flavonoids tends to bind more easily with the heteroatoms of amino acids from RBD/ACE2, as will be shown later. In this sense, peppermint flavonoids can compose the list of potential phytochemical inhibitors for the RBD/ACE2 interaction.  

Figures \ref{fig:protein_adsorption-01} and \ref{fig:protein_adsorption-02} illustrate the binding site surface (BSS) for the putative best docking target/ligand configurations. For the sake of clarity, these figures show the BSS only for the RBD/ACE2 region highlighted by the yellow rectangle in Figure \ref{fig:main_protein}(b). The following color scheme is adopted for the BSSs: gray, red, blue, and white for carbon, oxygen, nitrogen, and hydrogen atoms, respectively. In the ball-stick representation for the flavonoids, the carbon, oxygen, and hydrogen atoms are shown in the colors cyan, red, and white, respectively. As a general trend, one can note that the flavonoids fit inside the core pocket region (cavity) of the RBD/ACE2 complex. This cavity is displayed as region 2 in Figure \ref{fig:main_protein}(a). Acacetin, Luteolin$^{*}$, Luteolin$^{**}$, Thymusin, and Xanthomicrol were adsorbed on region 1 (see Figure \ref{fig:main_protein}(a)) of the RBD/ACE2 complex. The ligands tend to interact with the oxygen atoms (red spots in the BSS) in regions 1 and 2. These regions establish pocket-like media that yield interactions (mostly hydrogen bonds) between flavonoids and amino acid residues of proteins. 

Figures \ref{fig:protein-ligand-01} and \ref{fig:protein-ligand-02} provide a clear picture of the interaction between the amino acid residues of the proteins and peppermint flavonoids. The docked poses (obtained using PLIP \cite{salentin2015plip} show the residues names and the bond types. In the stick representation of flavonoids, the carbon and oxygen atoms are in the orange and red colors, respectively. The hydrogen, hydrophobic, and $\pi$-staking bonds are denoted by the blue, dashed gray, and dashed yellow lines, respectively. The yellow sphere represents the charge center. In Figure \ref{fig:protein-ligand-01} one can note that Acacetin, Apigenin, Apigenin$^{*}$, Chryseoriol, Hesperidin, Hesperetin, Ladenein, and Luteolin interact with RBD/ACE2 mainly through 4, 5, 5, 6, 12, 5, 4, and 8 hydrogen bonds with distinct amino acid residues in both RBD and ACE2 proteins. Similarly, Figure \ref{fig:protein-ligand-02} shows the interaction mechanism between Luteloin$^{*}$, Luteloin$^{**}$, Luteloin$^{***}$, Naringenin, Pebrellin, Sakuranetin, Thymusin, and Xanthomicrol with RBD/ACE2 is mediated by 7, 5, 9, 8, 5, 5, 4, and 4 hydrogen bonds with distinct amino acid residues in both RBD and ACE2 proteins, respectively. In total, 12 hydrophobic bonds were found. The flavonoids and amino acid residues of the proteins involved in this kind of interaction are highlighted below. Some $\pi$-stacking bonds are also present in the RBD/ACE2 interactions with flavonoids expecting for the Hesperidin (Figure \ref{fig:protein-ligand-01}(e)), Luteolin$^{*}$ (Figure \ref{fig:protein-ligand-02}(a)), and Xanthomicrol (Figure \ref{fig:protein-ligand-02}(h)) cases. 

Generally speaking, we identified 31 distinct amino acid residues of the RBD/ACE2 interacting with the peppermint flavonoids. The RBD amino acid residues (and their occurrence) are TYR738 (4), LYS682 (5), GLU761 (6), GLN674 (6), TYR770 (6), ARG688 (8), ASP670 (2), GLY761 (4), GLY741 (2), GLN39 (1), ALA740 (1), LYS723 (3), ARG673 (1), and SER759 (1). The ACE2 amino acid residues (and their occurrence) are GLU5 (3), SER1 (5), ASP12 (7), PHE372 (4), ARG375 (9), ASN15 (8), GLU19 (9), PRO371 (1), ANS15 (1), THR71 (1), ALA369 (4), ARG37 (1), ALA368 (1), LYS335 (2), ASP20 (1), TYR760 (1), and LYS8 (1). This result suggests that the target RBD/ACE2 amino acid residues for this class of phytochemicals are ARG375, ASN15, and GLU19 from ACE2, and ARG668 from RBD, based on their higher occurrence. The flavonoids that present hydrophobic bonds with the RBD/ACE2 amino acids, highlighted in the following as (flavonoid/residue), are Ladanein/GLU19, Luteolin/LYS682, Hesperetin/ASN15, Hesperetin/GLU19, Pebrellin/TYR760, Sakuranetin/GLU19, Thymusin/LY58, Acacetin/GLU5, Apigenin/ASN15, Apigenin/PRO371, Apigenin/TYR770, and Chryseoriol/LYS682.       

\section{Conclusions}

In summary, a set of phytochemicals (peppermint flavonoids) were screened against the SARS-CoV-2 Spike receptor-binding domain interacting with the human ACE2 receptor. The approach is based on computationally fitting small molecules for the target RBD/ACE2 complex proteins using the 3D structure of the active site with SWISSDOCK \cite{grosdidier2011fast,grosdidier2011swissdock}, subsequently the ranking of the docked compounds with Quimera \cite{pettersen2004ucsf} and interaction analysis with PLIP \cite{salentin2015plip}. Results revealed that Luteolin 7-O-neohesperidoside has a binding affinity of about -9.18 Kcal/mol, the higher one among the flavonoids studied here. On the other hand, Sakuranetin was the one with the lowest affinity (about -6.38 Kcal/mol). Binding affinities of the other peppermint flavonoids ranged from -6.44 Kcal/mol up to -9.05 Kcal/mol. These values outperform the binding affinities reported by docking studies using other types of compounds in which the RBD/ACE2 complex was also the target \cite{cherrak2020potential,bhowmik2021evaluation}. 

The binding site surface analysis showed pocket-like regions on the RBD/ACE2 complex that yield several interactions (mostly hydrogen bonds) between the flavonoid and the amino acid residues of the proteins. The interaction mechanism between the flavonoids and amino acid residues of the proteins is mediated by hydrogen bonds, essentially. The presence of some hydrophobic and $\pi-stacking$ bonds was also observed. In total, we identified 31 distinct amino acid residues of the RBD/ACE2 interacting with the peppermint flavonoids. The target RBD/ACE2 amino acid residues for this class of phytochemicals are ARG375, ASN15, and GLU19 from ACE2, and ARG668 from RBD, based on their higher occurrence. 

Some \textit{in vitro} studies investigated the antiviral activity of flavonoids in combating SARS-CoV \cite{badshah2021antiviral,RUSSO2020109211} and SARS-CoV2 \cite{abian2020structural,liskova2021flavonoids,seadawy2021vitro,solnier2020flavonoids} infection. Hesperetin, Luteolin, and Apigenin have been demonstrated as potent inhibitors of SARS-CoV-2 3CLpro \textit{in vitro} and can be considered proper candidates for further optimization and development of therapeutic interventions, particularly those related to inflammation processes and immunity \cite{solnier2020flavonoids}. A Luteolin derivative and Apigenin showed the best docking scores in our study.

\begin{acknowledgement}
The authors gratefully acknowledge the financial support from Brazilian Research Councils CNPq, CAPES, and FAPDF and CENAPAD-SP for providing the computational facilities. M.L.P.J. gratefully acknowledge the financial support from CAPES grant 88882.383674/2019-01. L.A.R.J. gratefully acknowledges respectively, the financial support from FAP-DF grant 00193.0000248/2019-32, CNPq grant 302236/2018-0, and UnB/DPI/DEX Edital 01/2020 grant 23106.057604/2020-05. The molecular graphics and analyses were performed with UCSF Chimera, developed by the Resource for Biocomputing, Visualization, and Informatics at the University of California, San Francisco, with support from NIH P41-GM103311.
\end{acknowledgement}

\bibliographystyle{spphys}
\bibliography{references}

\end{document}